\documentclass[twocolumn]{aa}

\usepackage{amsmath}
\usepackage[varg]{txfonts}
\usepackage{graphicx}
\usepackage[pdftex]{color, epsfig}
\usepackage{natbib}

\DeclareMathOperator\erf{erf}

\newcommand{\sm}[1]{\rm{{\scriptsize #1}}}
\newcommand{\simle} {\,{}^<_{\sim}\,}

\newcommand{\be}{\begin{equation}}
\newcommand{\ee}{\end{equation}}

\newcommand{\bea}{\begin{eqnarray}}
\newcommand{\eea}{\end{eqnarray}}

\newcommand{\SFE}{\rm{SFE}}
\newcommand{\MJ}{M_{\sm{J}}}
\newcommand{\NJ}{N_{\sm{J}}}

\newcommand{\Ma}{{\cal M}}
\newcommand{\Ca}{{\cal C}}

\newcommand{\dd}{{\mbox{d}}}

\newcommand{\bef}{\begin{figure}[!t]}
\newcommand{\eef}{\end{figure}}

\begin{document}

\title{Star formation efficiency in turbulent clouds}
\titlerunning{SFE in turbulent clouds}
\author{Robi Banerjee}
\authorrunning{R. Banerjee}
\institute{Hamburger Sternwarte, Universit\"at Hamburg, Gojenbergsweg
  112, 21029 Hamburg}
\offprints{{\tt banerjee@hs.uni-hamburg.de}}
\date{}

\abstract{Here we present a simple, but nevertheless, instructive
  model for the star formation efficiency ($\SFE$) in turbulent
  molecular clouds. The model is based on the assumption of log-normal
  density distribution which reflects the turbulent nature of the
  interstellar medium (ISM). Together with the number count of cloud
  cores, which follows a Salpeter-like core mass function (CMF), and the
  minimum mass for the collapse of individual cloud cores, given by
  the local Jeans mass ($M_{\sm{J}}$), we are able to derive the SFE
  for clouds as a function of their Jeans masses. We find a very
  generic power-law, $\SFE \propto \NJ^{-0.26}$, where $\NJ =
  M_{\sm{cloud}}/\MJ$ and a maximum $\SFE_{\sm{max}} \sim 1/3$ for the
  Salpeter case. This result is independent of the turbulent Mach
  number but fairly sensitive to variations of the CMF.}


\maketitle

\section{Introduction}
\label{sec:intro}

Molecular clouds, the birthplaces of stars in galaxies, are pervaded
by turbulent motions, which to a large extend determine the cloud's
density distribution \citep[see e.g. reviews by][and references
herein]{MacLow04, Ballesteros07, DobbsPPVI13, PadoanPPVI13}. The
distribution function of those density fluctuations is commonly
described by a log-normal distribution \citep[e.g.,][]{Vazquez94,
  Padoan02, Federrath08}. Furthermore, the mass distribution of cores
and clumps~\footnote{We use terms cores and clumps interchangeably for
  connected subregions within the considered cloud.}, i.e. the core
mass function (CMF), within the molecular cloud seems to follow a
power-law distribution, similar to the stellar initial mass function
(IMF) \citep[see e.g.,][]{Alves07, Rathborne09, Andre10, Koenyves10,
  Andre12}.  Based on those ingredients there are a number of analytic
approaches to calculate the efficiency of turbulent molecular clouds
to form stars. In particular, \citet{Padoan95} used a procedure
combining the distribution of cores according to the turbulent
property of the parent cloud with the mass distribution of those
cores. We comment again on the \citet{Padoan95} approach later in this
letter. \citet{KrumholzMcKee05} derived an efficiency of star
formation per free-fall time by considering the fraction of mass which
exceeds a critical density determined by the condition of
gravitational instability without taking the CMF into account, because
they are only interested in the rate at which stars form. Similar
approaches, i.e. considering the mass fraction of the density-PDF that
exceeds a critical density, were also used by e.g. \citet{Padoan11,
  HennebelleChabrier11, Kainulainen14}, to derive a star formation
efficiency.

In this letter, we would argue that it is not sufficient to identify
the mass of the high density fluctuations to calculate the star
formation efficiency, because, first this gives only a {\em lower}
mass-limit and second those high-density cores are embedded in lower
mean-density clumps which might still be able to collapse.  Therefore,
we derive an upper limit for the star formation efficiency by
explicitly calculating the highest-mass core within the entire cloud
which is able to collapse by gravitational instability.

\section{Model description}
\label{sec:model}

We start with the canonical form of the probability distribution function (PDF) of density fluctuations in a turbulent cloud,
\be
p(s) = \frac{1}{\sqrt{2\pi\,\sigma^2}}\,
\exp\left(-\frac{(s-s_0)^2}{2\,\sigma^2}\right)
\label{eq:PDF}
\ee
with $s = \ln(\rho/\rho_0)$ and $s_0 = -1/2\,\sigma^2$.  The variance
of this PDF depends on the turbulence as $\sigma^2 = \ln(1
+b^2\,\Ma^2)$, where for simplicity we neglect the impact of magnetic
fields \citep[see e.g.,][]{Vazquez94, Padoan02, Federrath08}. The
parameter $b$ is related to the type of turbulence \citep[see
again][]{Federrath08} but does not play a crucial role in our
consideration as we will see later on.

Obviously only those cloud cores will collapse and form stars which exceed the Jeans mass
\be
M_{\sm{J}} \approx \left(\frac{c_{\sm{s}}}{\sqrt{G}}\right)^3 \frac{1}{\sqrt{\rho}}
\propto \rho^{-1/2}
\ee
at their mean density $\rho$ and temperature $T \propto c_{\sm{s}}^2$ ($c_{\sm{s}}$ is the speed of sound and $G$ the gravitational constant). 

Using the mass of the cloud that exceeds the density $\rho$
\bea
M(s) & = & M_{\sm{cloud}}\, \int_s^{\infty}\dd s\, p(s)
\label{eq:M_of_s} \\
& = & \frac{ M_{\sm{cloud}}}{2}\, \left[1- \erf{\left(-\frac{s-s_0}{\sqrt{2\,\sigma^2}}\right)} \right]\nonumber
\eea
we can calculate the minimum mass $M_{\sm{min}}$ of a turbulent cloud which is Jeans unstable, i.e.
\be
M_{\sm{min}} : M(s) = M_{\sm{J}}(s) \, .
\ee

Given the fact that the cloud is fragmented by the same nature of
turbulence, the mass at a given density is not located in single cloud
cores but rather distrubuted according to a {\it core mass function}
(CMF). Interestingly, the CMF has a very similar shape the {\em
  stellar} initial mass function (IMF) and is often assumed to follow
a Salpeter distribution, i.e.
\be
{\rm CMF} \equiv \frac{\dd N}{\dd \ln{M}} \propto M^{-\alpha}
\ee
with $\alpha \approx 1.35$ \citep{Alves07}. Now the key point is the normalisation of this
number distribution which differs from cloud to cloud. The
normalisation of the CMF is given by the total mass of the cloud:
\be
M_{\sm{cloud}} = \Ca\,\int_{M_{\sm{low}}}^{M_{\sm{cloud}}}\dd M\, M^{-\alpha} \, .
\ee
It follows that ($\alpha \ne 1$)
\be
M_{\sm{cloud}} = \Ca\,\frac{M_{\sm{low}}^{-\alpha+1}}{\alpha-1}
\label{eq:norm}
\ee
assuming that $M_{\sm{cloud}} \gg M_{\sm{low}}$. Unfortunately on
first sight, this result depends strongly on the mass of the cores,
$M_{\sm{low}}$ which still contribute to the (Salpeter)
distribution. But fortunately, this mass can easily be determined as
we expect that the CMF is largely governed by the impact of
self-gravitating cloud cores \citep[e.g.,][]{Kainulainen11,
  Kainulainen13, Girichidis14}
\footnote{This is particular plausible if we assume that the IMF and
  CMF have a similar origin.}. In this case the lowest-mass core which
gives rise to the CMF is given by $M_{\sm{min}}$, i.e. the core which
is just Jeans unstable. Now the number distribution of the cores within
the cloud which masses are larger than $M$ is given by
\be
N(M) = \Ca\,\int_{M}^{M_{\sm{cloud}}}\dd M \, M^{-\alpha-1}
\ee
with $\Ca = (\alpha-1)\,(M_{\sm{cloud}}/M_{\sm{min}})\,M_{\sm{min}}^{\alpha}$ from Eq.~(\ref{eq:norm}). Hence,
\be
N(M) = \frac{\alpha-1}{\alpha}\,
 \left(\frac{M_{\sm{cloud}}}{M_{\sm{min}}}\right)\,
\left[
\left(\frac{M}{M_{\sm{min}}}\right)^{-\alpha}
- \left(\frac{M_{\sm{cloud}}}{M_{\sm{min}}}\right)^{-\alpha}
\right]
\label{eq:N_of_M}
\ee

Now we can search for the largest (locally connected) cloud core which is found by the condition
\be
M_{\sm{thres}} : N(M_{\sm{thres}}) = 1 \, .
\label{eq:sf_cond}
\ee

This condition tells us that there is only {\em one} core with mass
$M_{\sm{thres}}$, whereas cores that exceed this mass do {\em not}
exist in the cloud, i.e. $N(M) < 1$. Hence, cores with $M_{\sm{core}}
> M_{\sm{thres}}$ can {\em not} contribute to star formation, solely
because they are {\em not present}. Otherwise, cores that are smaller
than $M_{\sm{thres}}$ become increasingly more abundant for decreasing
core masses (as long as $\alpha > -1$). That means one could, in
principle, determine the smallest cores that might contribute to star
formation by $M(s)/N(s) > M_{\sm{J}}(s)$. But this conditions does
ignore that high density cores (which are the ones with the smallest
masses) are embedded in larger, more massive cores which are able to
form stars \citep{Vazquez94}.
Therefore, this condition does not apply for our consideration of the
SFE.

At this point we briefly have to comment on a similar approach
discussed by \citet{Padoan95}, \citep[see also][]{Padoan02}. Here the
efficiency to from stars is assumed to be essentially $M(m)\times
N(M)$ \citep[see Eqs. (21) and (24) of][]{Padoan95}, i.e. by the total
mass $M$ of {\em all} clumps with mass $m$ in the {\em entire} cloud
{\em times} the frequency of those clumps in the cloud. Hence, the
outcome of this convolution does not reflect the total mass of
unstable cores. 

Now we can calculate an upper limit for the SFE of molecular clouds
(MCs) and giant molecular clouds (GMCs) as a function of their number
of Jeans masses,
\be
{\rm SFE} \equiv \frac{M_{\sm{thres}}}{M_{\sm{cloud}}} \, ,
\label{eq:SFE_def}
\ee
where we use the number of Jeans masses
\be
\NJ \equiv \frac{M_{\sm{cloud}}}{\MJ(\rho_0)}
\ee
to quantify the instability of the cloud. 

\bef
  \centering
  \leavevmode
  \includegraphics[width=\linewidth]{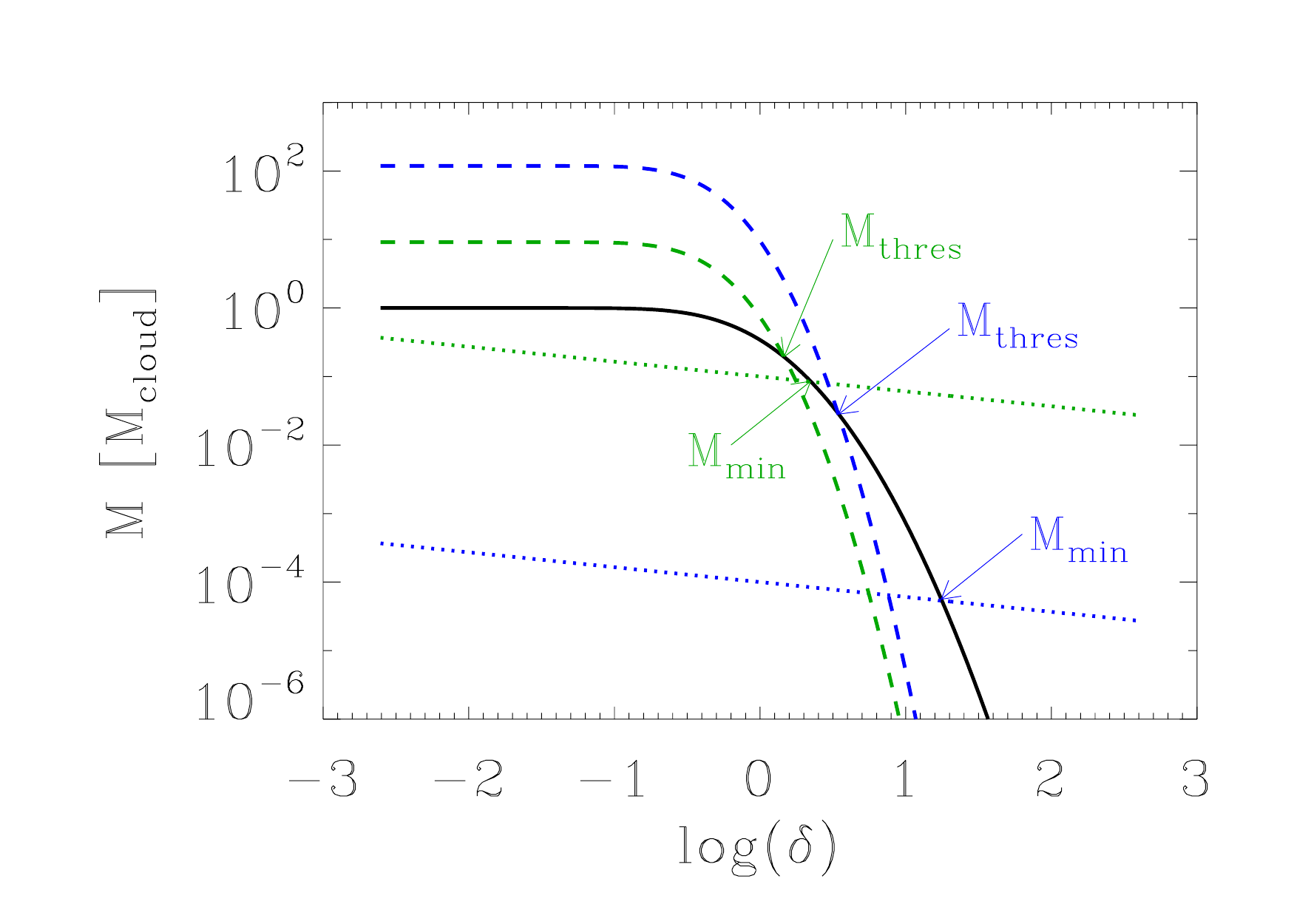}
  \caption{Shows the concept of the presented model on the $\SFE$ and density threshold for star formation. The intersection of the Jeans mass (dotted lines) with the mass distribution of the cloud (solid line) gives us the smallest  cores (by mass) which still are able to collapse. This minimum mass determins the upper end of the CMF. Calculating the individual core masses within the cloud (dashed lines) we can determine the largest core (by mass) which is present in the cloud (intersections with the solid line). Its mass is given by $M_{\sm{thres}}$ (see Eq.~\ref{eq:sf_cond}) and the ratio $M_{\sm{thres}}/M_{\sm{cloud}}$ gives us the uppper limit for the $\SFE$. Here we show two examples with $\NJ = 10$ (green lines) and $\NJ = 10^4$ (blue lines). \label{fig:masses}}
\eef
In Fig.~\ref{fig:masses} we summarise the concept of our model to calculate the $\SFE$ based on two examples of $\NJ$, where we use $\delta \equiv \rho/\rho_0$ 
for convenience.

\section{Results}
\label{sec:results}

Together with Eq.~(\ref{eq:N_of_M}) and the condition Eq.~(\ref{eq:SFE_def}) the star formation efficiency can be expressed as
\bea
\SFE & = & \left[ 
  \left(\frac{\alpha}{\alpha-1}\right)\, 
  \left(\frac{M_{\sm{cloud}}}{M_{\sm{min}}}\right)^{-1} 
+ \left(\frac{M_{\sm{cloud}}}{M_{\sm{min}}}\right)^{-\alpha}
\right]^{-1/\alpha}  \,
  \left(\frac{M_{\sm{cloud}}}{M_{\sm{min}}}\right)^{-1} \nonumber \\
& \approx & \left(\frac {\alpha-1}{\alpha}\right)^{1/\alpha}
  \left(\frac{M_{\sm{cloud}}}{M_{\sm{min}}}\right)^{(1-\alpha)/\alpha} \, ,
\label{eq:SFE}
\eea
where we assumed $\alpha > 1$ for the approximation. Measuring the cloud mass in terms of its Jeans mass at the mean density we see from Eq.~(\ref{eq:SFE}) that
\be
\SFE \propto \NJ^{(1-\alpha)/\alpha}
\ee
which results in $\SFE \propto \NJ^{-0.26}$ for the Salpeter case.

Obviously, the definition Eq.~(\ref{eq:SFE_def}) is the largest value
a cloud could achive if all the mass of unstable cores are converted
instantaneously into stars ignoring all kinds of additional effects
like feedback from the stars themself. But this picture incorporates
the effect of reduced accretion onto stars by fragmentation of the
cloud, i.e. this model quantifies the consequence of {\em
  Fragmentation Induced Starvation} (FIS) \citep{Peters10c,
  Girichidis12a}, which can be seen from Fig~\ref{fig:SFE_NJ}. The more
unstable the cloud, quantified by $\NJ$,
the less efficient it can from stars because it is more prone to
fragmentation with a number of fragments which are not Jeans unstable
anymore. Interestingly, in the Salpeter case ($\alpha = 1.35$) the
maximal $\SFE$ is $\sim 1/3$~\footnote{the more precise number is
  $36.6\%$} for clouds with $M_{\sm{cloud}} \approx M_{\sm{J}}$. This
applies, for instance, to isolated Bok globules like Barnard 68 which
might be barely unstable \citep{Alves01}. Even such low-mass clouds
could only convert at most $\sim 1/3$ of their mass into stars, even
without any feedback, because they will fragment while they are
collapsing. 

\bef
  \centering
  \leavevmode
  \includegraphics[width=.9\linewidth]{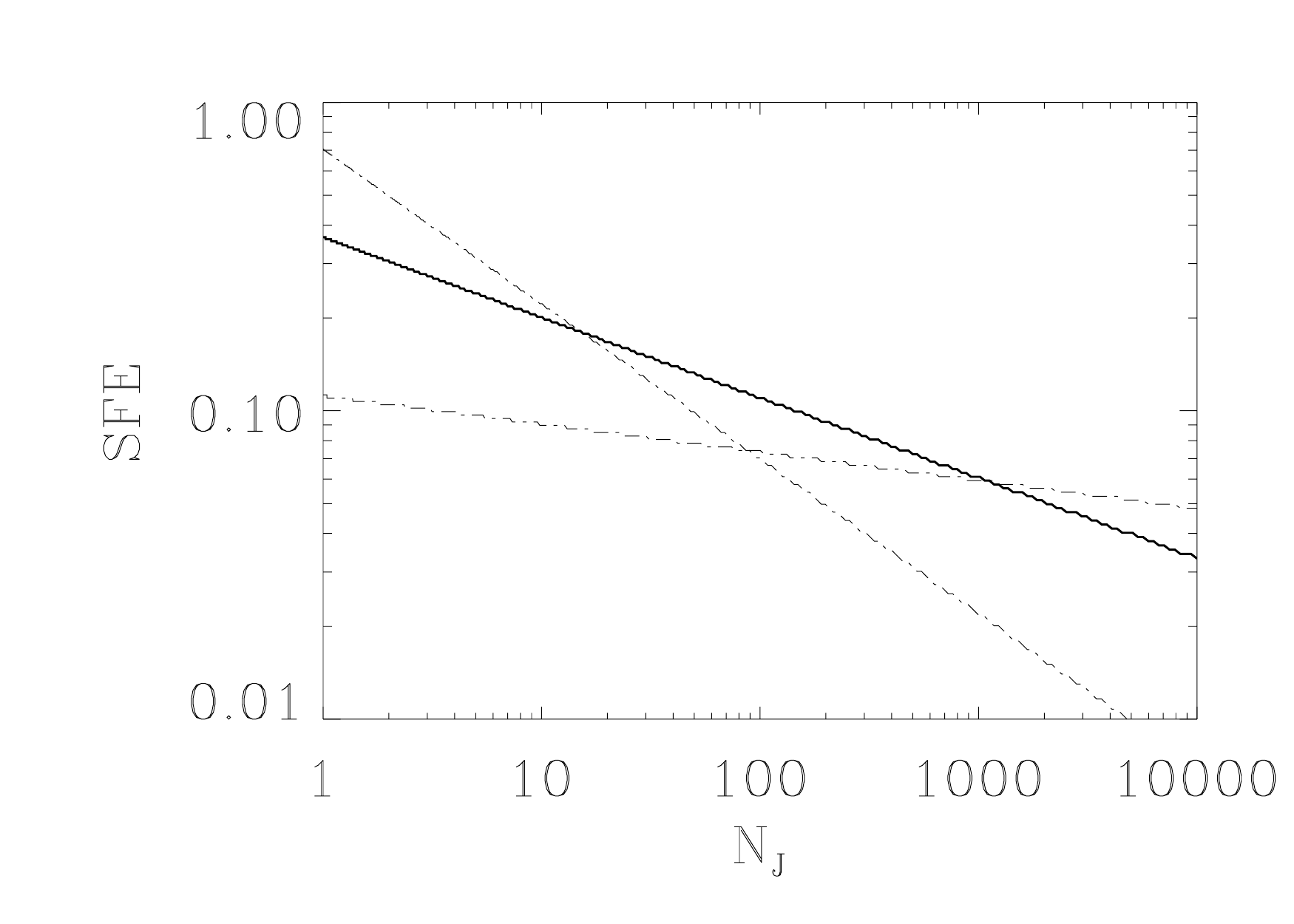}
  \caption{$\SFE$ for our fiducial model, i.e. $\Ma = 1$, $\alpha = 1.35$ (solid line), and for $\alpha=2$ and $\alpha=1.1$ (upper and lower dashed line, respectively). \label{fig:SFE_NJ}}
\eef

\bef
  \centering
  \leavevmode
  \includegraphics[width=.9\linewidth]{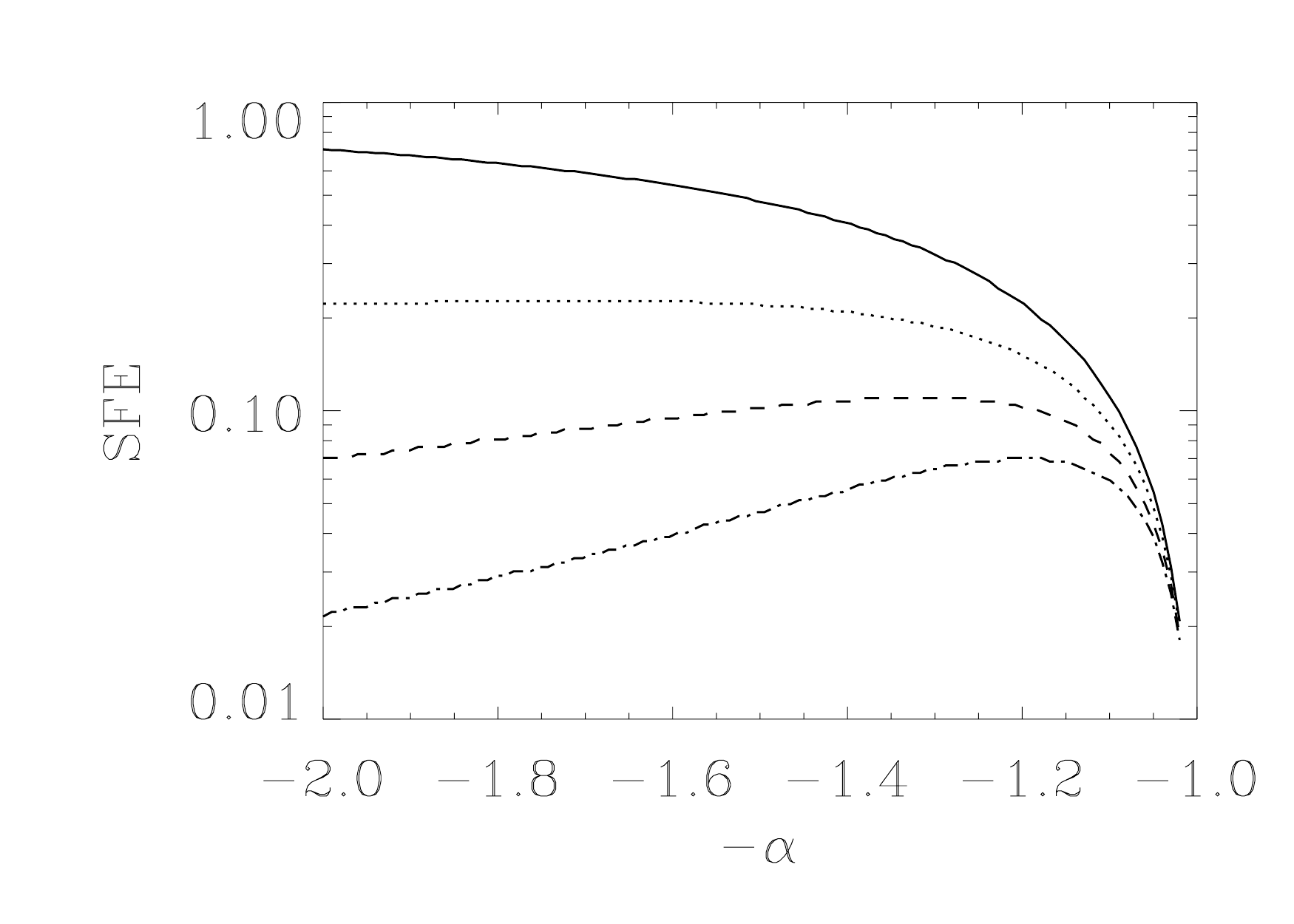}
  \caption{$\SFE$ as a function of the CMF-slope $\alpha$ for
    different instability parameters. The lines from top to bottom are
    for $\NJ = 1,10,100$ and $1000$, respectively. Up to $\NJ \simle
    10$ there is a clear decresing trend of the SFE with decresing
    concentration of the cloud. For massive clouds, this trend is
    reversed up to a minimal value of $\alpha =
    \alpha_{\sm{min}}(\NJ)$ (see text). \label{fig:SFE_alpha}}
\eef

Also interesting is the fact that the $\SFE$ decreases with decreasing
concentration of the cloud (decreasing $\alpha$) for less unstable
systems ($\NJ \simle 10$). Again the reason is the fragmentation
property of the cloud. Less concentrated clouds are more susceptible
to fragmentation than clouds with a high density concentration. This
behaviour is intensively studied in \citet{Girichidis11a} and
\citet{Girichidis12a} where the collaps of clouds with various density
profiles were investigated. Nevertheless, the situations gets a bit
more complicated for more unstable clouds ($\NJ > 10$) as seen in
Fig.~\ref{fig:SFE_alpha}. Here, less concentration of the CMF helps to
{\em increase} the $\SFE$ up to a certain maximal value depending on
$\NJ$ and $\alpha$. Hence, for more unstable clouds, the enhanced
fragmentation helps to a certain degree to increase the SFE as such
fragments are still Jeans unstable and therefore contribute to star
formation. This competition between constructive fragmentation and
rapid collapse is only efficient up to a minimal concentration,
$\alpha_{\sm{min}}$ of the cloud (e.g., for $\NJ = 1000 \rightarrow
\alpha_{\sm{min}} \approx 1.2$). For less concentrated clouds, $\alpha
\simle 1.1$, the SFE becomes essentially independent of $\NJ$ and
approches zero in the limiting case $\alpha \rightarrow 1$.


\bef
  \centering
  \leavevmode
  \includegraphics[width=.9\linewidth]{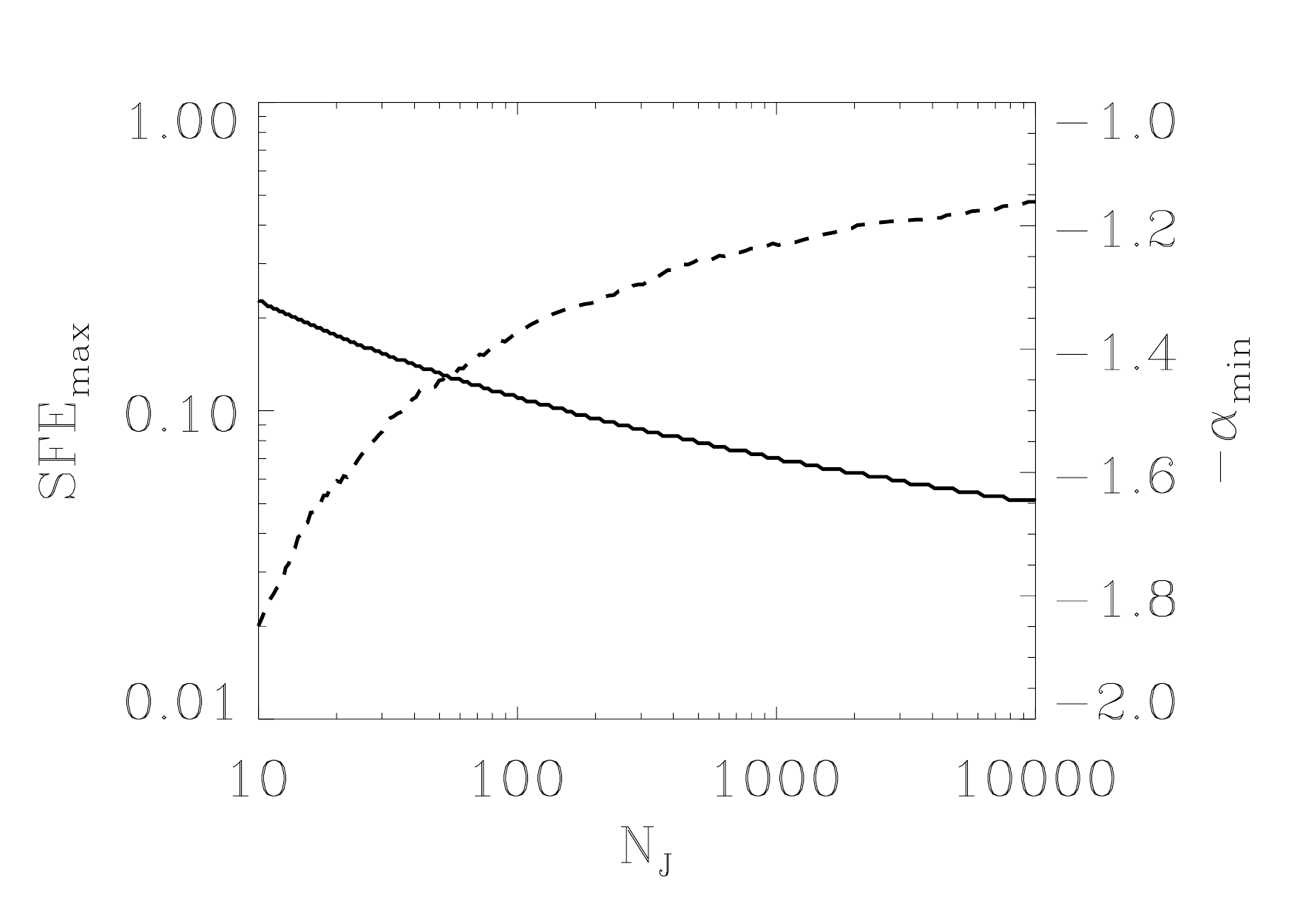}
  \caption{The maximal star formation efficiency, $\SFE_{\sm{max}}$
    (solid line, left axis) and the appropriate CMF-slope
    $\alpha_{\sm{min}}$ (dashed line, right axis) as a function of the
    number of Jeans masses, $\NJ$. These values are determined by the
    fact that for rather unstable clouds the $\SFE$ as a function of
    $\alpha$ reaches a maximum at $\alpha_{\sm{min}}$ (see
    Fig.~\ref{fig:SFE_alpha} and text). \label{fig:SFE_max}}
\eef

Another interesting aspect which one obtains from the above
consideration of the threshold mass, $M_{\sm{thres}}$, is the
threshold density, $\rho_{\sm{thres}}$, for the onset of star
formation~\footnote{Actually, the discussion with Marcel V\"olschow on
  a self-consistent description of a SF-density threshold spawned this
  project.}. Going back to the mass distribution $M(s)$ given by
Eq.~(\ref{eq:M_of_s}), one can read off $\rho_{\sm{thres}}$ from the
solution of Eq.~(\ref{eq:sf_cond}).

\bef
  \centering
  \leavevmode
  \includegraphics[width=.9\linewidth]{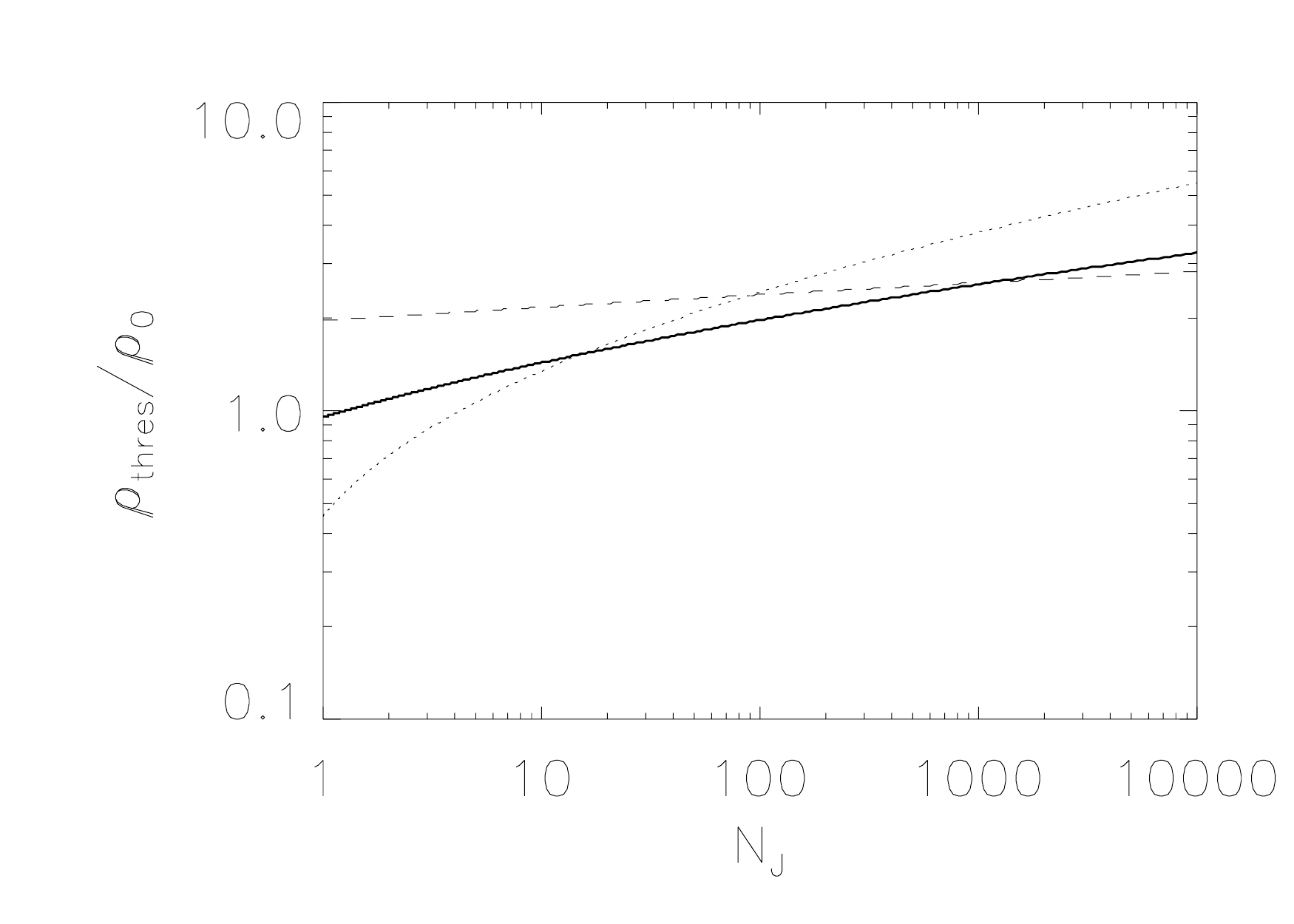}
  \caption{The threshold density for star formation for $\alpha =
    1.35$ (solid line), $\alpha = 1.1$ (dashed line) and $\alpha = 2$
    (dotted line). Similar to the $\SFE$, only for $\NJ \simle 10$
    there is a trend of a decreasing threshold with increasing CMF
    concentration (see also text). 
\label{fig:dens_thres}}
\eef

We present the result for different values of $\alpha$ in
Fig.~\ref{fig:dens_thres}. First of all we see, that there is very
litte dependence of $\rho_{\sm{thres}}$ on the instability of the
cloud (in the $\alpha = 1.1$ case it becomes almost independent of
$\NJ$) and $\rho_{\sm{thres}}/\rho_0$ is of order unity. This is not
too surprising as we only consider globally unstable clouds in the
first place. Again, only for $\NJ \simle 10$ we find a clear trend of
$\rho_{\sm{thres}}$ with the cloud concentration $\alpha$: Less
concentrated clouds need a larger threshold density to produce stars
compared to those with a steeper CMF. For more unstable clouds, $\NJ
> 10$, the competition between fragmentation and collapse does
not lead to such a clear trend with the cloud concentration. 

\paragraph{Mach number dependence}

So far we presented our results for transonic molecular clouds with
$\Ma =1$ (assuming $b=1$, see also Eq.~(\ref{eq:PDF}) and below). But
it turns out that neither the type of turbulence nor its strength has
a large impact on our results. This can already be seen from
Eq.~(\ref{eq:SFE}) which has only a very weak dependence on
$M_{\sm{min}}$, i.e. on the quantity which depends on $\Ma$. We tested
the Mach number dependence of the $\SFE$ numerically using the basic
equations, but could not see any visible difference. Hence, we omit a
plot showing $\SFE$ as a function of $\Ma$. 
The weak dependency of $\SFE$ on $\Ma$ can be understood as
follows: Larger Mach numbers result in wider distributions of the
density-PDF and therefore would give rise to a larger density
threshold (or smaller $M_{\sm{min}}$) for the same Jeans mass. But a
wider PDF also reduces the overall instability of the cloud,
i.e. reduces $\NJ$. Both effects almost compensate each other. But
please note, that already the dependency of $M_{\sm{min}}$ on $\Ma$
is very weak as can be seen from Fig.~\ref{fig:masses}.

\section{Conclusions}
\label{sec:conclusions}

Here we presented a simple model for the star formation efficiency in
turbulent molecular clouds. The model is based on the assumption of
log-normal density distribution which reflects the turbulent nature of
the ISM. Similar to previous analytic studies, we use this
distribution to estimate the minimum mass which can actually collapse
by gravitational instability. Any cores that mass exceeds
$M_{\sm{min}}$ are also Jeans-unstable, but, according to the density
distribution, have a lower mean density and are less frequent than
lower-mass cores. The latter statement reflects the observed
distribution of core masses. But following the CMF, not all
low-density regions exist as connected clumps, and hence are not
Jeans-unstable. Combining the density-PDF and the CMF we calculate
largest core within the cloud which is still able to collapse. This in
turn can be used to infer an upper limit for the $\SFE$.  For a given
slope of the CMF, we find a very generic power-law, $\SFE \propto
\NJ^{-(\alpha-1)/\alpha}$ and a maximum $\SFE_{\sm{max}} \approx 0.37$
for the Salpeter case. Again, this result is independent of the
turbulent Mach number.

\begin{acknowledgements} 
I thank Marcel V\"olschow for inspiring discussions that initiated
this letter. I am grateful to the Deutsche Forschungsgemeinschaft
(DFG)  for funding the projects BA 3706/1-1, BA 3706/3-1, BA 3706/3-2
and BA 3706/4-1.

\end{acknowledgements}


\begin{thebibliography}{24}
\expandafter\ifx\csname natexlab\endcsname\relax\def\natexlab#1{#1}\fi

\bibitem[{{Alves} {et~al.}(2007){Alves}, {Lombardi}, \& {Lada}}]{Alves07}
{Alves}, J., {Lombardi}, M., \& {Lada}, C.~J. 2007, \aap, 462, L17

\bibitem[{{Alves} {et~al.}(2001){Alves}, {Lada}, \& {Lada}}]{Alves01}
{Alves}, J.~F., {Lada}, C.~J., \& {Lada}, E.~A. 2001, \nat, 409, 159

\bibitem[{{Andr{\'e}} {et~al.}(2010){Andr{\'e}}, {Men'shchikov}, {Bontemps},
  {K{\"o}nyves}, {Motte}, {Schneider}, {Didelon}, {Minier}, {Saraceno},
  {Ward-Thompson}, {di Francesco}, {White}, {Molinari}, {Testi}, {Abergel},
  {Griffin}, {Henning}, {Royer}, {Mer{\'{\i}}n}, {Vavrek}, {Attard},
  {Arzoumanian}, {Wilson}, {Ade}, {Aussel}, {Baluteau}, {Benedettini},
  {Bernard}, {Blommaert}, {Cambr{\'e}sy}, {Cox}, {di Giorgio}, {Hargrave},
  {Hennemann}, {Huang}, {Kirk}, {Krause}, {Launhardt}, {Leeks}, {Le Pennec},
  {Li}, {Martin}, {Maury}, {Olofsson}, {Omont}, {Peretto}, {Pezzuto}, {Prusti},
  {Roussel}, {Russeil}, {Sauvage}, {Sibthorpe}, {Sicilia-Aguilar}, {Spinoglio},
  {Waelkens}, {Woodcraft}, \& {Zavagno}}]{Andre10}
{Andr{\'e}}, P., {Men'shchikov}, A., {Bontemps}, S., {et~al.} 2010, \aap, 518,
  L102

\bibitem[{{Andr{\'e}} {et~al.}(2012){Andr{\'e}}, {Men'shchikov}, {K{\"o}nyves},
  {Schneider}, {Arzoumanian}, {Peretto}, {Palmeirim}, {Didelon},
  {Ward-Thompson}, {Di Francesco}, {Bontemps}, {Motte}, {Molinari}, \&
  {Herschel Gould Belt Consortium}}]{Andre12}
{Andr{\'e}}, P., {Men'shchikov}, A., {K{\"o}nyves}, V., {et~al.} 2012, in From
  Atoms to Pebbles: Herschel's view of Star and Planet Formation

\bibitem[{{Ballesteros-Paredes} {et~al.}(2007){Ballesteros-Paredes}, {Klessen},
  {Mac Low}, \& {Vazquez-Semadeni}}]{Ballesteros07}
{Ballesteros-Paredes}, J., {Klessen}, R.~S., {Mac Low}, M.-M., \&
  {Vazquez-Semadeni}, E. 2007, in Protostars and Planets V, ed. B.~{Reipurth},
  D.~{Jewitt}, \& K.~{Keil}, 63--80

\bibitem[{{Dobbs} {et~al.}(2013){Dobbs}, {Krumholz}, {Ballesteros-Paredes},
  {Bolatto}, {Fukui}, {Heyer}, {Mac Low}, {Ostriker}, \&
  {V{\'a}zquez-Semadeni}}]{DobbsPPVI13}
{Dobbs}, C.~L., {Krumholz}, M.~R., {Ballesteros-Paredes}, J., {et~al.} 2013,
  ArXiv e-prints

\bibitem[{{Federrath} {et~al.}(2008){Federrath}, {Klessen}, \&
  {Schmidt}}]{Federrath08}
{Federrath}, C., {Klessen}, R.~S., \& {Schmidt}, W. 2008, \apjl, 688, L79

\bibitem[{{Girichidis} {et~al.}(2011){Girichidis}, {Federrath}, {Banerjee}, \&
  {Klessen}}]{Girichidis11a}
{Girichidis}, P., {Federrath}, C., {Banerjee}, R., \& {Klessen}, R.~S. 2011,
  \mnras, 413, 2741

\bibitem[{{Girichidis} {et~al.}(2012){Girichidis}, {Federrath}, {Banerjee}, \&
  {Klessen}}]{Girichidis12a}
{Girichidis}, P., {Federrath}, C., {Banerjee}, R., \& {Klessen}, R.~S. 2012,
  \mnras, 420, 613

\bibitem[{{Girichidis} {et~al.}(2014){Girichidis}, {Konstandin}, {Whitworth},
  \& {Klessen}}]{Girichidis14}
{Girichidis}, P., {Konstandin}, L., {Whitworth}, A.~P., \& {Klessen}, R.~S.
  2014, \apj, 781, 91

\bibitem[{{Hennebelle} \& {Chabrier}(2011)}]{HennebelleChabrier11}
{Hennebelle}, P. \& {Chabrier}, G. 2011, \apjl, 743, L29

\bibitem[{{Kainulainen} {et~al.}(2011){Kainulainen}, {Beuther}, {Banerjee},
  {Federrath}, \& {Henning}}]{Kainulainen11}
{Kainulainen}, J., {Beuther}, H., {Banerjee}, R., {Federrath}, C., \&
  {Henning}, T. 2011, \aap, 530, A64

\bibitem[{{Kainulainen} {et~al.}(2014){Kainulainen}, {Federrath}, \&
  {Henning}}]{Kainulainen14}
{Kainulainen}, J., {Federrath}, C., \& {Henning}, T. 2014, Science, 344, 183

\bibitem[{{Kainulainen} \& {Tan}(2013)}]{Kainulainen13}
{Kainulainen}, J. \& {Tan}, J.~C. 2013, \aap, 549, A53

\bibitem[{{K{\"o}nyves} {et~al.}(2010){K{\"o}nyves}, {Andr{\'e}},
  {Men'shchikov}, {Schneider}, {Arzoumanian}, {Bontemps}, {Attard}, {Motte},
  {Didelon}, {Maury}, {Abergel}, {Ali}, {Baluteau}, {Bernard}, {Cambr{\'e}sy},
  {Cox}, {di Francesco}, {di Giorgio}, {Griffin}, {Hargrave}, {Huang}, {Kirk},
  {Li}, {Martin}, {Minier}, {Molinari}, {Olofsson}, {Pezzuto}, {Russeil},
  {Roussel}, {Saraceno}, {Sauvage}, {Sibthorpe}, {Spinoglio}, {Testi},
  {Ward-Thompson}, {White}, {Wilson}, {Woodcraft}, \& {Zavagno}}]{Koenyves10}
{K{\"o}nyves}, V., {Andr{\'e}}, P., {Men'shchikov}, A., {et~al.} 2010, \aap,
  518, L106

\bibitem[{{Krumholz} \& {McKee}(2005)}]{KrumholzMcKee05}
{Krumholz}, M.~R. \& {McKee}, C.~F. 2005, \apj, 630, 250

\bibitem[{{Mac Low} \& {Klessen}(2004)}]{MacLow04}
{Mac Low}, M.-M. \& {Klessen}, R.~S. 2004, Reviews of Modern Physics, 76, 125

\bibitem[{{Padoan}(1995)}]{Padoan95}
{Padoan}, P. 1995, \mnras, 277, 377

\bibitem[{{Padoan} {et~al.}(2013){Padoan}, {Federrath}, {Chabrier}, {Evans},
  {Johnstone}, {J{\o}rgensen}, {McKee}, \& {Nordlund}}]{PadoanPPVI13}
{Padoan}, P., {Federrath}, C., {Chabrier}, G., {et~al.} 2013, ArXiv e-prints

\bibitem[{{Padoan} \& {Nordlund}(2002)}]{Padoan02}
{Padoan}, P. \& {Nordlund}, {\AA}. 2002, \apj, 576, 870

\bibitem[{{Padoan} \& {Nordlund}(2011)}]{Padoan11}
{Padoan}, P. \& {Nordlund}, {\AA}. 2011, \apj, 730, 40

\bibitem[{{Peters} {et~al.}(2010){Peters}, {Klessen}, {Mac Low}, \&
  {Banerjee}}]{Peters10c}
{Peters}, T., {Klessen}, R.~S., {Mac Low}, M.-M., \& {Banerjee}, R. 2010, \apj,
  725, 134

\bibitem[{{Rathborne} {et~al.}(2009){Rathborne}, {Lada}, {Muench}, {Alves},
  {Kainulainen}, \& {Lombardi}}]{Rathborne09}
{Rathborne}, J.~M., {Lada}, C.~J., {Muench}, A.~A., {et~al.} 2009, \apj, 699,
  742

\bibitem[{{Vazquez-Semadeni}(1994)}]{Vazquez94}
{Vazquez-Semadeni}, E. 1994, \apj, 423, 681

\end{thebibliography}

\end{document}